\documentclass[aip,reprint]{revtex4-1}
\usepackage{graphicx}
\usepackage{dcolumn}
\usepackage{bm}
\usepackage{amsmath}
\usepackage{textcomp}
\def\Xint#1{\mathchoice
{\XXint\displaystyle\textstyle{#1}} 
{\XXint\textstyle\scriptstyle{#1}} 
{\XXint\scriptstyle\scriptscriptstyle{#1}} 
{\XXint\scriptscriptstyle\scriptscriptstyle{#1}} 
\!\int}
\def\XXint#1#2#3{{\setbox0=\hbox{$#1{#2#3}{\int}$ }
\vcenter{\hbox{$#2#3$ }}\kern-.59\wd0}}

\def\dashint{\Xint-}
\begin{document}
\title{Capillary-driven flow induced by a stepped perturbation atop a viscous film} 
\author{Thomas Salez}\email{thomas.salez@gmail.com}
\address{Laboratoire de Physico-Chimie Th\'eorique, UMR CNRS Gulliver 7083, ESPCI, Paris, France}
\author{Joshua D. McGraw}
\address{Department of Physics \& Astronomy and the Brockhouse Institute for Materials Research, McMaster University, Hamilton, Canada}
\author{Oliver B\"{a}umchen}
\address{Department of Physics \& Astronomy and the Brockhouse Institute for Materials Research, McMaster University, Hamilton, Canada}
\author{Kari Dalnoki-Veress}
\address{Department of Physics \& Astronomy and the Brockhouse Institute for Materials Research, McMaster University, Hamilton, Canada}
\author{Elie Rapha\"{e}l}
\address{Laboratoire de Physico-Chimie Th\'eorique, UMR CNRS Gulliver 7083, ESPCI, Paris, France}
\date{\today}

\begin{abstract}
Thin viscous liquid films driven by capillarity are well described in the lubrication theory through the thin film equation. In this article, we present an analytical solution of this equation for a particular initial profile: a stepped perturbation. This initial condition allows a linearization of the problem making it amenable to Fourier analysis. The solution is obtained and characterized. As for a temperature step in the heat equation, self-similarity of the first kind of the full evolution is demonstrated and a long-term expression for the excess free energy is derived. In addition, hydrodynamical fields are described. The solution is then compared to experimental profiles from a model system: a polystyrene nanostep above the glass transition temperature which flows due to capillarity. The excellent agreement enables a precise measurement of the capillary velocity for this polymeric liquid, without involving any numerical simulation. More generally, as these results hold for any viscous system driven by capillarity, the present solution may provide a useful tool in hydrodynamics of thin viscous films.
\end{abstract}

\maketitle

\section*{Introduction}
Micro- and nanofilms are of tremendous importance in a variety of scientific fields \cite{Oron1997,Craster2009,Blossey2012}, such as polymer physics, physiology, biophysics, micro-electronics, surface chemistry, thermodynamics or hydrodynamics. For instance, they are involved in modern mechanical and optical engineering processes, through lubrication, paints and coating. Gaining a complete understanding of these systems is a key step towards the development of molecular electronics, biomimetics, superadhesion and self-cleaning surfaces. 

Thin films furthermore remain of fundamental interest in physics and mathematics, as they raise important questions that are still unsolved. The example of polymer systems, which we will preferentially refer to throughout the present article, is enlightening on this very point. As far as rheology of ultra-thin polymer films is concerned, when the height of the film is comparable to the characteristic size of the macromolecule, scaling arguments have been proposed for the effective viscosity \cite{Brochard2000}. Several of the recent experimental results that have been obtained motivate the desire for a fundamental understanding of polymers in confinement. Examples are the enhancement of the effective mobility in the liquid \cite{Bodiguel2006} and glassy state \cite{Fakhraai2008}, and long-term relaxation of bulk viscosity \cite{Barbero2009}. The modification of  polymeric conformations \cite{Jones1999} and interchain entanglements \cite{Si2005,Baumchen2009} near surfaces, and in confined geometries \cite{Shin2007}, have been studied in detail. Surface instabilities and pattern formation have been explored as well \cite{Mukherjee2011,Closa2011,Amarandei2012}. Moreover, the role of film preparation has been investigated \cite{Stillwagon1990,Raegen2010} but remains complex, as pointed out in the particular case of the glass transition temperature \cite{Reiter2001,Baumchen2012}.

The systems discussed above are often well described by the lubrication theory through capillary-driven thin film equations \cite{Oron1997,Craster2009}. However, due to their high orders and non-linearities, these particular equations have not yet been solved analytically. Nevertheless, the reader will find details on the mathematical advances in recent reviews \cite{Myers1998,Kondic2003}. In addition, we point out the fact that thin film equations have been solved numerically in various configurations \cite{Bertozzi1998,Salez2012a}. 

\begin{figure}
\includegraphics[width=7cm,trim = 0mm 2mm 4.5mm 8mm, clip=true]{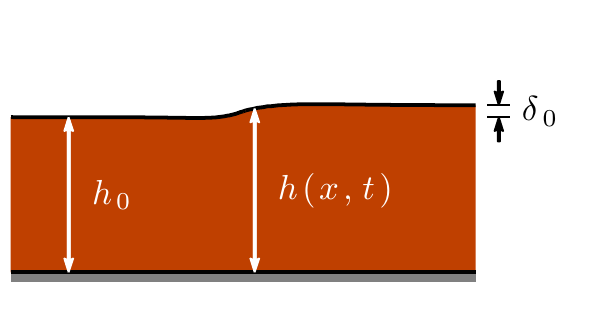}
\caption{\label{fig:scheme} \textit{A schematic showing the capillary leveling of an initially stepped perturbation with amplitude $\delta_0$ atop a thin viscous film of height $h_0$. The vertical profile $h(x,t)$ of the free surface depends only on the horizontal coordinate $x$ and time $t$.}}
\end{figure}

In the present communication, we address the analytical problem of a \textit{stepped perturbation} atop a flat film (shown schematically in Fig.~\ref{fig:scheme}) which is infinite in the two horizontal dimensions. The interest in this system is twofold. First, it is physically relevant since it is inspired by experiment \cite{McGraw2011, McGraw2012}. Secondly, it is mathematically interesting since the smallness of the  step in comparison to the underlying film height allows one to linearize the thin film equation and solve it through Fourier analysis, as for the heat equation \cite{Fourier1822}.

This study is divided into three sections. In the first one, we recall the main ingredients of the physical model, including the general non-linear thin film equation for two-dimensional capillary-driven flows. In the second part, we explicitly consider the stepped perturbation, for which we obtain and characterize the linear solution. Finally, in the third part, we compare the theoretical results to experiments performed on a polystyrene (PS) \textit{stepped film} above the glass transition temperature.

\section{Physical framework}
In this first section, we describe the physical model that is considered. After justifying the main assumptions and giving the boundary conditions for the flow, we derive the general non-linear thin film equation in two dimensions.  
 
\subsection{Assumptions}
\label{assumptions}
We consider the example of PS nanofilms above their glass transition temperature $T_{\textrm{g}}\sim100\ ^{\circ}\textrm{C}$, since this experimental system will be compared to the theory in section~\ref{exp}. We thus have the following typical parameters \cite{Wu1970,Rubinstein2003,Brandrup2005}: vertical height $h_0\sim1$~\textmu m, dynamical viscosity $\eta\sim1\ \textrm{MPa.s}$, molecular weight $M_\textrm{w}\sim15$~kg.mol$^{-1}$, surface tension $\gamma\sim30\ \textrm{mN.m}^{-1}$, density $\rho\sim1\ \textrm{g.cm}^{-3}$ and shear modulus $G\sim1$~MPa. Extension to any thin viscous fluid is straightforward by using the corresponding relevant orders of magnitude. 

Let us first estimate the typical spreading capillary velocity $v_\textrm{c}$, Reynolds number $\textrm{Re}$, capillary length $l_{\textrm{c}}$ and Maxwell viscoelastic time $\tau_{\textrm{M}}$:
\begin{subequations}
\begin{align}
v_\textrm{c}&=\frac{\gamma}{\eta}\sim2\ \textrm{\textmu m.min}^{-1}\\
\textrm{Re}&=\frac{h_0\rho v_\textrm{c}}{\eta}\ll 1\\
l_{\textrm{c}}&=\sqrt{\frac{\gamma}{\rho g}}\sim 2\ \textrm{mm}\gg h\\
\tau_{\textrm{M}}&=\frac{\eta}{G}\sim1\ \textrm{s}\ .
\end{align}
\end{subequations}
From these orders of magnitude, and since we typically observe slow evolution of the profiles over several minutes \cite{McGraw2011,McGraw2012}, we can make the following assumptions: we have an incompressible flow of a highly viscous Newtonian fluid, where gravity \cite{Huppert1982}, disjoining pressure \cite{Seemann2001} and inertia are negligible. Therefore, this flow is well described by the Stokes equation:
\begin{equation}
\label{stokes}
\mathbf{\nabla}P=\eta\mathbf{\Delta v}\ ,
\end{equation}
combined with the incompressibility condition:
\begin{equation}
\label{div}
\mathbf{\nabla\cdot v}=0\ ,
\end{equation}
where $P$ and $\mathbf{v}$ are the local pressure and velocity fields in the liquid. In addition, we assume that the profile slopes remain small in comparison to $1$, which is an ingredient of the lubrication approximation. Finally, we take $\gamma$ and $\eta$ as homogeneous and constant. Note that the case of inhomogeneous surface tension is considered elsewhere \cite{Oron1997,Craster2009} and that we neglect confinement effects\cite{Brochard2000,Bodiguel2006,Fakhraai2008} since the film height is still large in comparison with the size of the macromolecule, typically $\sim 10$~nm for the considered molecular weight. 

\subsection{Vertical boundary conditions}
\label{bc}
We consider the case of no shear at the liquid-air interface:
\begin{equation}
\label{shear}
\partial_z \mathbf{v} |_{z=h}=\mathbf{0}\ ,
\end{equation}
where $z$ is the vertical coordinate. In addition, we assume no slip at the substrate:
\begin{equation}
\label{slip}
v_{\parallel} |_{z=0}=0\ ,
\end{equation}
with $v_{\parallel}$ the component of the velocity parallel to the liquid-substrate interface.

\subsection{Thin film equation}
In addition to the assumptions and the vertical boundary conditions presented in the previous parts, we assume a spatial invariance along one horizontal direction $y$. We then have a pure two-dimensional problem. Therefore, the height of the free surface is given by $h(x,t)$, where $x$ is the relevant horizontal direction and $t$ the time. The local pressure field is \textit{a priori} given by $P(x,z,t)$. According to the lubrication approximation, we can neglect the vertical velocity with respect to the horizontal one and write: $\mathbf{v}=v(x,z,t)\ \mathbf{e_x}$, where $\mathbf{e_x}$ is the horizontal basis vector. Then, we project and integrate Eq.~(\ref{stokes}), using Eq.~(\ref{div}), Eq.~(\ref{shear}) and Eq.~(\ref{slip}), and we find:
\begin{equation}
\label{press}
\partial_z P=0\ ,
\end{equation}
the local pressure field $P(x,t)$ is thus invariant in the vertical direction, and:
\begin{equation}
\label{poiseuille}
v(x,z,t)=\frac{1}{2\eta}(z^2-2hz)\ \partial_x P\ ,
\end{equation} 
which corresponds to the usual parabolic Poiseuille flow. Volume conservation requires that:
\begin{equation}
\label{isov}
\partial_t h+\partial_x\int_0^{h}dz\ v=0\ .
\end{equation} 
Finally, because the pressure does not depend on $z$ (see Eq.~(\ref{press})) it can be evaluated at the free surface through the Young-Laplace equation. Since the lubrication approximation implies small curvatures, the pressure satisfies:
\begin{equation}
\label{laplace}
P-P_0\approx-\gamma\partial_x^{\,2}h\ ,
\end{equation} 
where $P_0$ is the atmospheric pressure. Thus, combining Eq.~(\ref{poiseuille}), Eq.~(\ref{isov}) and Eq.~(\ref{laplace}), we get:
\begin{equation}
\label{tfe}
\partial_th+\frac{\gamma}{3\eta}\partial_x\left(h^3\partial_x^{\,3}h\right)=0\ ,
\end{equation} 
which is known \cite{Stillwagon1988,Stillwagon1990,Oron1997,Bertozzi1998,Craster2009} as the capillary-driven thin film equation.

\section{Stepped perturbation}
\label{secltfe}
In the present section, we consider an infinitesimal perturbation $\delta(x,t)\ll h_0$ of the free surface on an infinitely large and flat thin film with height $h_0$ (see Fig.~\ref{fig:scheme}). We thus linearize the thin film equation, before nondimensionalizing the problem and giving its general formal solution through Fourier analysis. Then, we consider a simple initial profile: a stepped film, for which we express and characterize the solution. In particular, we demonstrate self-similarity of the first kind of the full evolution and we study the long-term viscous dissipation of the excess free energy, before describing the hydrodynamical fields of the problem.

\subsection{Linearized Thin Film Equation}
In the following, we restrict ourselves to the linearized version of the thin film equation. Setting:
\begin{equation}
\label{defdel}
h(x,t)=h_0+\delta(x,t)\ , 
\end{equation}
Eq.~(\ref{tfe}) becomes:
\begin{equation}
\label{ltfe}
\partial_t\delta+\frac{\gamma h_0^{\,3}}{3\eta}\partial_x^{\,4}\delta=0\ ,
\end{equation} 
to first order in the perturbation. Apart from the higher order in the spatial derivative, this equation is analogous to the heat equation \cite{Fourier1822}. 

\subsection{Non-dimensionalizing}
For generality, we introduce the typical height $h_0$ and time $t_0=3\eta h_0/\gamma$ of the problem, as well as the associated dimensionless variables:
\begin{subequations}
\label{natl}
\begin{align}\label{natlength}
\Delta&=
\frac{\delta}{h_0}\\
\label{natlength2}
X&=
\frac{x}{h_0}\\
\label{natlength3}
T&=
\frac{t}{t_0}\ .
\end{align} \end{subequations}
We non-dimensionalize Eq.~(\ref{tfe}) and obtain:
\begin{equation}
\label{adltfe}
\partial_T\Delta+\partial_X^{\ 4}\Delta=0\ .
\end{equation}

\subsection{Formal general solution}
Injecting a propagating mode $\textrm{e}^{i(KX-\Omega T)}$, of angular frequency $\Omega$ and angular wavenumber $K$, into Eq.~(\ref{adltfe}) leads to the dispersion equation:
\begin{equation}
\label{disp}
\Omega=-iK^4\ ,
\end{equation}
which implies in particular that spatial oscillations with small wavelengths decay in amplitude faster than those with large wavelengths. Since Eq.~(\ref{adltfe}) is linear, we can write the general solution as a superposition in the Fourier basis. Using Eq.~(\ref{disp}), it follows that:
\begin{equation}
\label{sum}
\Delta(X,T)=\int_{-\infty}^{+\infty}\frac{dK}{\sqrt{2\pi}}\ A(K)\ \mathrm{e}^{-K^4T}\mathrm{e}^{\mathrm{i}KX} \ ,
\end{equation}
where $A(K)$ is the spatial Fourier transform of the initial profile:
\begin{equation}
\label{coef}
A(K)=\int_{-\infty}^{+\infty}\frac{dX}{\sqrt{2\pi}}\ \Delta(X,0)\ \mathrm{e}^{-\mathrm{i}KX}\ .
\end{equation}

\subsection{Solution for a stepped initial condition}
We consider now the particular ideal case where the initial profile is a step proportional to the Heaviside function $\Theta$:
\begin{equation}
\label{ci}
\Delta(X,0)=\Delta_0\ \Theta(X)\ ,
\end{equation}
through a dimensionless amplitude $\Delta_0=\delta_0/h_0$.

Under the initial condition of Eq.~(\ref{ci}), Eq.~(\ref{sum}) becomes:
\begin{equation}
\label{sol}
\Delta(X,T)=\frac{\Delta_0}{2}\left[1+\psi(X,T)\right] \ ,
\end{equation}
where we shifted the vertical origin and normalized the amplitude through the centered profile:
\begin{equation}
\label{sol1}
\psi(X,T)=\dashint_{-\infty}^{+\infty}dK\ \frac{\mathrm{e}^{-K^4T}}{\mathrm{i}\pi K}\mathrm{e}^{\mathrm{i}KX}\ ,
\end{equation}
and where the dashed integral represents Cauchy's principal value. 

\subsection{Limits and symmetry}
First, Eq.~(\ref{sol1}) naturally reduces to the sign function at $T=0$, as required from the initial condition through Eq.~(\ref{ci}). Secondly, the energy cost associated with the surface perturbation is viscously dissipated and the final equilibrium state is thus flat at finite $X$:
\begin{equation}
\label{fs}
\lim\limits_{T \to +\infty}\Delta(X,T)=\frac{\Delta_0}{2}\ .
\end{equation}
Thirdly, at finite $T$, using Cauchy's residue theorem and Jordan's lemma, and provided we can perform an inversion of limits,  we obtain:
\begin{subequations}
\label{sss}
\begin{align}
\lim\limits_{X \to +\infty}\Delta(X,T)&=\Delta_0\\
\lim\limits_{X \to -\infty}\Delta(X,T)&=0\ .
\end{align} \end{subequations}
This important result tells us that the horizontal boundary limits remain those of the initial profile all along the evolution, which stresses the fact that the boundary limits of the initial profile are crucial in defining the symmetry of the shape. Moreover, the leveling does not contaminate the infinite horizontal limits at finite time, as expected.
 
Finally, there is a fixed point at $X=0$, along the temporal evolution:
\begin{equation}
\label{fp}
\Delta(0,T)=\frac{\Delta_0}{2} \ ,
\end{equation}
which is a center of symmetry because $\psi$ is an odd function with respect to $X$.

\subsection{Self-similarity}
At finite $T$, let us change variables through:
\begin{subequations}
\begin{align}\label{ss1}
X&=UT^{1/4}\\
K&=\frac{Q}{T^{1/4}}\ .
\end{align}\label{ssstep} \end{subequations}
Thus, Eq.~(\ref{sol1}) becomes:
\begin{subequations}
\begin{align}\label{ss2}
\psi(X,T)&=
\chi(U)\\
\label{sol2}
&=\dashint_{-\infty}^{+\infty}dQ\ \frac{\mathrm{e}^{-Q^4}}{\mathrm{i}\pi Q}\mathrm{e}^{\mathrm{i}QU}\ .
\end{align} \end{subequations}
Therefore, with the initial profile considered, the full evolution is self-similar of the first kind \cite{Barenblatt1996}: it depends only on the variable $U$ and the profile satisfies the following equation \cite{Aradian2001}:
\begin{equation}
\chi''''=\frac{U}{4}\chi'\ ,
\end{equation}
whose integral has been studied in the mathematical literature \cite{Pfeiffer1972a,Pfeiffer1972b}. Once again, this equation is analogous to the one obtained with self-similar solutions in $XT^{-1/2}$ for the heat equation \cite{Fourier1822}, apart from the order of the derivative on the left-hand side. Note that the self-similarity introduced in Eq.~(\ref{ssstep}) and Eq.~(\ref{ss2}) is highly dependent on Eq.~(\ref{sss}), and thus on the boundary limits of the initial profile. Different boundary limits would lead to different forms of self-similarity \cite{Cormier2012}. In fact, we see through Eq.~(\ref{sum}) that the condition for such a self-similarity of the solution is that the Fourier transform of the initial profile is positive homogeneous of degree $-1$: 
\begin{equation}
A(\alpha K)=\alpha^{-1}A(K)\ ,
\end{equation}
for $\alpha>0$. In other words, according to Eq.~(\ref{coef}):
\begin{equation}
\Delta(\alpha X,0)=\Delta(X,0)\ .
\end{equation}
Therefore, the initial profile has to be a constant function with a potential discontinuity at $X=0$, \textit{i.e.} the self-similarity considered can only be achieved if there is a step initially, which justifies \textit{a posteriori} the interest of this configuration. In addition, note that these results are analogous to the ones obtained for the heat equation \cite{Fourier1822} with a temperature step, except for the precise time exponent which is related to the order of the spatial derivative in the governing equation. In this picture, we see that the final equilibrium state (see Eq.~(\ref{fs})) and fixed point (see Eq.~(\ref{fp})) values are unified into a single point of the self-similar profile:  
\begin{equation}
\chi(0)=0\ .
\end{equation}
Moreover, the profile is symmetric with respect to this particular central point:
\begin{equation}
\label{odd}
\chi(-U)=-\chi(U)\ .
\end{equation}

Finally, we see through Eq.~(\ref{sol2}) that $\chi$ is proportional to the inverse Fourier transform of the function $Q\mapsto Q^{-1}\textrm{e}^{-Q^4}$. Therefore, the solution is:
\begin{equation}
\label{sol3}
\chi(U)=\frac{2\Gamma(5/4)}{\pi}\ U\ \textrm{F}_{\alpha}(U)-\frac{\Gamma(3/4)}{12\pi}\ U^3\ \textrm{F}_{\beta}(U)\ ,
\end{equation}
where we introduced two auxiliary functions for clarity:
\begin{subequations}
\begin{align}\textrm{F}_{\alpha}(U) &=\, 
_1\textrm{F}_3\left(\left\{\frac{1}{4}\right\},\left\{\frac{2}{4},\frac{3}{4},\frac{5}{4}\right\},\left(\frac{U}{4}\right)^4\right)\\
\textrm{F}_{\beta}(U) &=\, 
_1\textrm{F}_3\left(\left\{\frac{3}{4}\right\},\left\{\frac{5}{4},\frac{6}{4},\frac{7}{4}\right\},\left(\frac{U}{4}\right)^4\right)\ ,
\end{align} \end{subequations}
with the definition of the $(1,3)$-generalized hypergeometric function \cite{Gradshteyn1965,Abramowitz1965}:
\begin{equation}
_1\textrm{F}_3\left(\{a\},\{b,c,d\},w\right)=\sum_{k\ge0}\frac{(a)_k}{(b)_k(c)_k(d)_k}\frac{w^k}{k!}\ ,
\end{equation}
and the Pochhammer symbol $(.)_k$ of the rising factorial. Note that the self-similar dimensionless solution in Eq.~(\ref{sol3}) is unique: it depends only on the variable $U$ and thus not on the experimental parameters $h_0$, $\delta_0$, $\gamma$, $\eta$ and $t$.

\begin{figure}
\includegraphics[width=8.6cm]{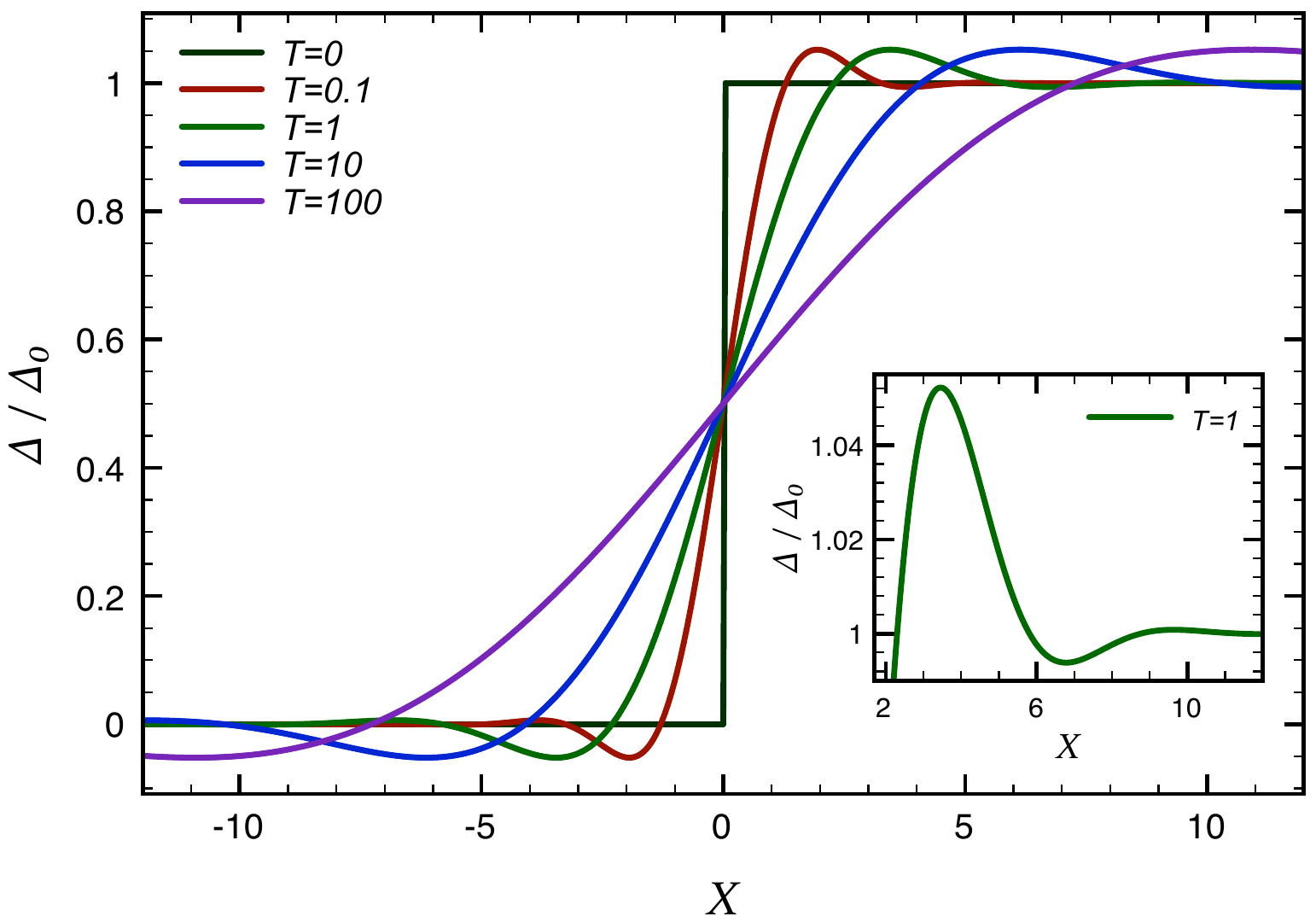}
\caption{\label{fig:step}\textit{Solution of Eq.~(\ref{adltfe}) for a stepped initial profile (see Eq.~(\ref{ci})) at various dimensionless times, according to Eq.~(\ref{sol}), Eq.~(\ref{ss1}), Eq.~(\ref{ss2}) and Eq.~(\ref{sol3}). The profiles are symmetric with respect to the central fixed point, according to Eq.~(\ref{odd}). The self-similar profile is naturally identical to the real profile at $T=1$, since $\chi(U)=\psi(U,1)$ according to Eq.~(\ref{ss1}) and Eq.~(\ref{ss2}). The inset shows a zoom of the first oscillations for $X>0$, in the $T=1$ profile.}}
\end{figure}

Fig.~\ref{fig:step} shows $\Delta/\Delta_0$ as a function of $X$ for various times.
As we can see, the profile flattens through time as the excess surface energy is viscously dissipated. Apart from the oscillatory behavior linked with the fourth spatial derivative of Eq.~(\ref{adltfe}), this evolution is qualitatively close to the solution of the heat equation \cite{Fourier1822} for which the same analytical treatment would lead to the usual error function. At fixed time $T$, the spatial oscillatory behavior (see inset of Fig.~\ref{fig:step}) is qualitatively analogous to the temporal response of a damped harmonic oscillator after having switched on an external constant load, although the spectrum is different according to Eq.~(\ref{sol1}). Finally, plotting the same data with respect to the self-similar variable $U$, instead of variable $X$ for a given parameter $T$, would allow to collapse all the data onto the $\Delta(U,1)$ profile since $\chi(U)=\psi(U,1)$, by construction. 

\subsection{Extrema}
Let us now consider the amplitude of the oscillatory behavior. A local extremum at $U_i^*$ of the self-similar profile is defined by $\chi'(U^*_i)=0$. So, we write:
\begin{equation}
\label{min}
\int_{-\infty}^{+\infty}dQ\ \mathrm{e}^{-Q^4}\mathrm{e}^{\mathrm{i}QU_i^*}=0\ .
\end{equation}
It is straightforward to see that $\chi'(-U^*_i)=0$ as well, and that $\chi''(-U^*_i)=-\chi''(U^*_i)$. Therefore, each local maximum (minimum) has a mirror minimum (maximum), whose horizontal coordinate is symmetric with respect to $U=0$. This is trivial when recalling the symmetry of the profile through Eq.~(\ref{odd}). Interestingly, Eq.~(\ref{min}) implies that $U^*_i$ is independent of $\Delta_0$: the amplitude of the perturbation only shifts the vertical origin and stretches vertically the self-similar function $\chi$.  This scaling invariance of the solution is not expected to hold for the general non-linear thin film equation, due to the cubic dependency on the profile height (see Eq.~(\ref{tfe})).

We now focus on two mirror extrema indexed by $i$ and whose coordinates are: $\left[U^*_i,\chi(U^*_i)\right]$ and $\left[-U^*_i,\chi(-U^*_i)\right]$, with $U^*_i>0$. By introducing the ratio:
\begin{equation}
\label{bd}
\mathcal{R}=\left |\frac{\chi(U^*_i)-\chi(+\infty)}{\chi(-U^*_i)-\chi(-\infty)}\right |\ ,
\end{equation}
and by invoking the symmetry of the profile through Eq.~(\ref{odd}), we naturally get:
\begin{equation}
\mathcal{R}=1\ ,
\end{equation}
which is indeed observed in Fig.~\ref{fig:step}. An important feature is that this ratio does not depend on the amplitude of the perturbation $\Delta_0$. This scaling invariance of the solution does not hold for the general non-linear thin film equation \cite{McGraw2012}, due to the cubic dependency on the profile height (see Eq.~(\ref{tfe})).

\subsection{Excess free energy}
In order to characterize the relaxation to equilibrium, we study the long-term evolution of the capillary free energy (per unit length along $y$) in excess of the equilibrium flat profile. At large time and thus small slopes, this quantity satisfies:
\begin{equation}
\label{energ}
\Delta\mathcal{F}\approx\frac{\gamma}{2}\int dx\ \left(\partial_x \delta\right)^2\ .
\end{equation}
Rewriting Eq.~(\ref{energ}) in dimensionless variables through Eq.~(\ref{natl}) gives:
\begin{equation}
\Delta\mathcal{F}=\frac{\gamma h_0}{2}\int dX\ \left(\partial_X \Delta\right)^2\ .
\end{equation}
Finally, invoking the variable $U$ from Eq.~(\ref{ss1}) leads to the desired long-term expression:
\begin{equation}
\frac{\Delta\mathcal{F}}{\Delta\mathcal{F}_0}=\xi\left(\frac{\tau}{t}\right)^{1/4}\ , 
\end{equation}
where we introduced the initial excess capillary energy $\Delta\mathcal{F}_0=\gamma\delta_0$, and the characteristic time:
\begin{equation}
\tau=\frac{\eta \delta_0^{\,4}}{\gamma h_0^{\,3}}\ ,
\end{equation} 
as well as a constant numerical factor:
\begin{equation}
\xi= \frac{3^{1/4}}{8}\int dU\ \chi'^2(U)\approx0.16\ .
\end{equation}
The final relation comes from a calculation of the integral $\int \chi'^2 \approx 1$, using Eq.~(\ref{sol3}).
Although the geometrical dependencies of this result are only limits of the non-linear case, the dependencies on material properties do hold in the non-linear case \cite{McGraw2012}. This important long-term result teaches us that the surface energy is viscously dissipated with a $1/4$ power-law in the inverse time, and that the initial geometry plays the crucial role of a driving amplitude. 

\subsection{Hydrodynamical fields}
Using the solution obtained in Eq.~(\ref{sol3}), we now derive the expressions of the pressure and velocity fields in real variables, as well as the corresponding volumetric flow rate. As for the previous treatment of energy, using Eq.~(\ref{poiseuille}), Eq.~(\ref{laplace}), Eq.~(\ref{defdel}), Eq.~(\ref{natl}), Eq.~(\ref{sol}), Eq.~(\ref{ss1}) and Eq.~(\ref{ss2}), we find:
\begin{equation}
P(x,t)= P_0-\chi''(U)\left(\frac{3\eta\gamma\delta_0^{\, 2}}{4h_0^{\,3}t}\right)^{1/2}\ ,
\end{equation}
and:
\begin{equation}
\label{velo}
v(x,z,t)\approx\frac{3^{3/4}}{4}\chi'''(U)\left(\frac{\gamma\delta_0^{\,4}}{\eta h_0t^3}\right)^{1/4}\left(\frac{2z}{h_0}-\frac{z^2}{h_0^{\, 2}}\right)\ ,
\end{equation}
to lowest order in the perturbation. In particular, the velocity at the central fixed point $x=0$ satisfies:
\begin{equation}
v_0(z,t)\approx-0.1\left(\frac{\gamma\delta_0^{\,4}}{\eta h_0t^3}\right)^{1/4}\left(\frac{2z}{h_0}-\frac{z^2}{h_0^{\, 2}}\right)\ ,
\end{equation}
where we calculated the numerical factor through Eq.~(\ref{sol3}). The minus sign is expected due to the chosen geometry (see Fig.~\ref{fig:scheme}). Interestingly, we see that the amplitude of the parabolic velocity profile, \textit{i.e.} the surface velocity, decreases with time, viscosity and film height, and that it increases with surface tension and amplitude of the perturbation. Then, by integrating Eq.~(\ref{velo}) over $z$, we obtain the volumetric flow rate (per unit length in the $y$-direction):
\begin{equation}
\mathcal{Q}(x,t)\approx\frac{1}{2}\frac{\chi'''(U)}{3^{1/4}}\left(\frac{\gamma\delta_0^{\,4}h_0^{\,3}}{\eta t^3}\right)^{1/4}\ ,
\end{equation}
to lowest order in the perturbation. Thus, at the central fixed point $x=0$, we have:
\begin{equation}
\mathcal{Q}_0(t)\approx-0.07\left(\frac{\gamma\delta_0^{\,4}h_0^{\,3}}{\eta t^3}\right)^{1/4}\ .
\end{equation}
The dependencies are the same as for the velocity field, except for the film height: when $h_0$ tends to infinity the flow diverges, as expected, whereas the velocity tends to zero. 

\section{Comparison with experiments}
\label{exp}
In this last section, we compare the analytical solution obtained in section~\ref{secltfe} to a model experiment \cite{McGraw2011, McGraw2012}: a PS stepped perturbation above the glass transition temperature (see Fig.~\ref{fig:scheme}). In particular, we fit the profiles to the solution of Eq.~(\ref{sol3}). As we finally see, this procedure enables an accurate measurement of the capillary velocity of the material.

\subsection{Self-similarity}
\begin{figure}
\includegraphics[width=8.7cm]{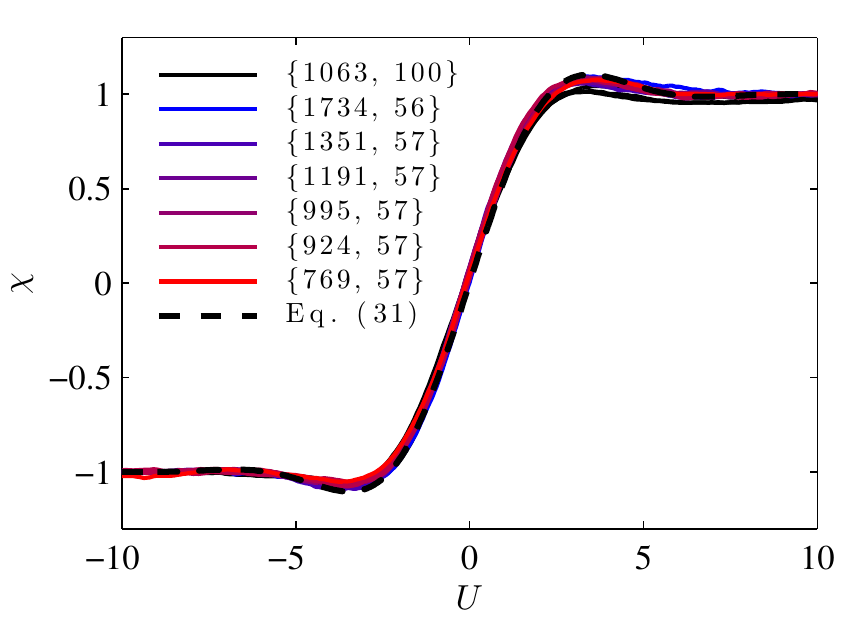}
\caption{\label{fit} \textit{Comparison between theory (see Eq.~(\ref{sol3})) and experiments (see Eq.~(\ref{ssexp})). The experimental profiles correspond to AFM lines taken at room temperature after quenching from $120^\circ\mathrm{C}$. PS stepped films were prepared as detailed elsewhere \cite{McGraw2011, McGraw2012}.  The experimental data represent steps with different heights $\{h_0,\delta_0\}$ in $~\mathrm{nm}$, as indicated. The profiles are recorded after $t=10~\mathrm{min}$ of annealing; exceptions are for: \{1063, 100\}, where there are three profiles corresponding to $t=6$, $8$ and $10~\mathrm{min}$; and for $\{924, 57\}$ where there are two profiles corresponding to $t =10$ and $20~\mathrm{min}$. There is only one free horizontal stretching parameter in this comparison, which leads to a measurement of the capillary velocity: $\gamma/\eta=1.5\pm0.1\ $}\textmu\textit{$\mathrm{m.min}^{-1}$, according to Eq.~(\ref{ufit}).}}
\end{figure}

Knowing the profile $\delta(x,t)$ and the film height $h_0$ for a given experimental sequence, it is straightforward to obtain the experimental equivalents (in real variables) of Eq.~(\ref{natlength}): 
\begin{equation}
\Delta^{\textrm{e}}(x,t)=\frac{\delta(x,t)}{h_0}\ ,
\end{equation}
and of Eq.~(\ref{sol}):
\begin{equation}
\label{psie}
\psi^{e}(x,t)=2\frac{\delta(x,t)}{\delta_0}-1\ .
\end{equation}  
We checked the self-similarity by plotting the experimental profiles as a function of the real self-similar variable \cite{McGraw2012}: 
\begin{equation}
\label{equ}
u=\frac{x}{t^{1/4}}\ .
\end{equation}
When doing so, the profiles at different times collapse onto a single profile:
\begin{equation}
\label{ssexp}
\psi^{e}(x,t)=\chi^{\textrm{e}}(u)\ , 
\end{equation}
thus demonstrating the self-similarity of the first kind of the experimental evolution. This result appears in Fig.~\ref{fit}.

\subsection{Fitting the profiles}
When comparing the profiles $\chi^{\textrm{e}}(u)$ from the experiment (see Eq.~(\ref{ssexp})) and $\chi(U)$ from the theory (see Eq.~(\ref{sol3})), we immediately see that a horizontal stretch factor remains due to the nondimensionalizing procedure. In fact, according to Eq.~(\ref{natl}), Eq.~(\ref{ss1}) and Eq.~(\ref{equ}), we have:
\begin{equation}
\label{ufit}
U=\left(\frac{3\eta}{\gamma h_0^{\,3}}\right)^{1/4}u\ .
\end{equation}
We thus fit all the experimental profiles $\chi^{\textrm{e}}(u)$ to the theoretical one $\chi(U)$ through this single free parameter. The result is plotted in Fig.~\ref{fit} for various $h_0$, $\delta_0$ and $t$.

The agreement is excellent and holds for height ratios up to $\delta_0/h_0\sim10\%$. In particular, we see that the experimental profiles plotted this way do not depend on $h_0$, $\delta_0$ and $t$, as suggested by the theory. The fact that the scaled profiles are independent of the stepped film geometry confirms the uniqueness and self-similarity of the dimensionless solution. Furthermore, since we know $h_0$ experimentally, this single parameter fitting procedure offers a precise measurement of the capillary velocity $\gamma/\eta$ of the material at the considered temperature. We find: $\gamma/\eta=1.5\pm0.1\ \textrm{\textmu m.min}^{-1}$ at $120^\circ\textrm{C}$, which compares well with the tabulated values \cite{Wu1970,Bach2003} through the WLF model \cite{Williams1955}.

As a final remark, we note that the initial stepped profile seems in contradiction with the lubrication approximation used in the theory. Thus, the short-term evolution of the experimental profile may not be described by the theoretical solution in Eq.~(\ref{sol3}). However, as observed experimentally, the step levels due to Laplace pressure gradients and reaches small slopes on a timescale that is short in comparison to the one of all the experimental observations. Furthermore, the experimental profiles connect rapidly to the theoretical solution. Therefore, the possible temporal offset is negligible in comparison with the typical experimental times and does not reduce the precision on the capillary velocity measurement.

\section*{Conclusion}
We reported on an analytical solution of the linear thin film equation for a stepped initial condition. The solution was obtained by performing Fourier analysis and involved generalized hypergeometric functions. We characterized the solution and demonstrated the self-similarity of the first kind of the full evolution. In addition, we found the origin of this symmetry in the initial profile itself. Then, using self-similarity, we derived the long-term viscous dissipation law in such a linear system. This scaling captures the relevant physical ingredients and provides a limit for the corresponding expression in the non-linear theory. We also derived the scaling expressions for the pressure and velocity fields and for the volumetric flow rate. As far as the linearized equation is concerned, apart from the well known case of the heat equation, we suspect that hypergeometric solutions may be obtained in an identical way for higher even orders of the spatial derivative in the linear equation. However, the odd orders being free from dissipation are expected to lead to fundamentally different mathematical solutions. Finally, we compared the solution to experimental profiles obtained on polystyrene stepped films for several times, and height ratios up to $\sim10\%$. The agreement is excellent, thus demonstrating the interest of such a solution in the physics of capillary-driven thin viscous films. In particular, the fitting technique offers a precise viscometer without invoking any numerical simulation. In the near future, these results may be extended to other initial conditions and to the first non-linear term of the equation through perturbation theory. The goals would be to study the convergence to intermediate self-similarity and to capture the non-linear extension of the present results in order to approach the full non-linear solution, which is still an open analytical problem that governs a considerable number of exciting physical applications.

\begin{acknowledgments}
The authors would like to thank Marco Fontelos, Anne-Laure Dalibard and Justin Salez for useful mathematical references. They thank as well as the \'Ecole Normale Sup\'{e}rieure of Paris, the Natural Sciences and Engineering Research Council of Canada, the German Research Foundation (DFG) under grant BA 3406/2, the Chaire Total-ESPCI and the Saint Gobain Fellowship for financial support. 
\end{acknowledgments}

\end{document}